\documentclass[a4paper]{jpconf}
\usepackage{graphicx}
\newtheorem{theorem}{Theorem}
\newtheorem{dex}{Definition}

\newcommand{\ket}[1]{|#1\rangle}

\begin{document}

\title{Condition for tripartite entanglement}%

\author{Allan I. Solomon}%
\address{Department of Physics and Astronomy, Open University, UK \\
 and LPTMC, University of Paris VI, France}
\ead{a.i.solomon@open.ac.uk}%
\author{Choon-Lin Ho}
\address{Department of Physics, Tamkang University, Tamsui
251, Taiwan, R.O.C.}
\ead{hcl@mail.tku.edu.tw}%

\date{14 Nov 2011} 

\begin{abstract}
We propose  a  scheme for classifying the entanglement of a
tripartite pure qubit state. This classification scheme  consists
of an ordered list of seven elements. These elements are the
Cayley hyper-determinant, and its six associated $2 \times 2$
subdeterminants.  In particular we show that this classification
provides a necessary and sufficient condition for separability.

\end{abstract}

\section{Introduction}

\medskip

It is well known that for bipartite pure qubit states a single
determinental condition is enough to discriminate between
separability and entanglement. It is a straightforward matter  to
determine whether a vector $v \in V = V_1\otimes V_2$ is entangled
or not. Here $V_1$ and $V_2$ are two-dimensional (qubit) vector
spaces with basis $\{e_1 \equiv \ket{0},e_2 \equiv \ket{1}\}$.

In general we may write  $v \in V$ as
\begin{equation}\label{bip}
  v \in V_1\otimes V_2 =\sum_{i,j=1}^{2}c_{ij} \; \; e_i \otimes e_j
\end{equation}
If $v$ is non-entangled, i.e. separable, then
\begin{equation}
 v= (x_{1}e_{1}+x_{2}e_{2})\otimes(y_{1}e_{1}+y_{2}e_{2})
\end{equation}
so
\begin{equation}\label{bisep}
   c_{ij}=x_{i}y_{j} \;\;\;\; \{i,j=1,2\}
\end{equation}
from which we deduce that the matrix $c$ of coefficients $c_{ij}$ has determinant zero, $\det c=0$.
And this condition is clearly necessary and sufficient.

In fact, by suitably normalizing, we may use this determinant to
provide a {\em measure} of entanglement for pure states called the
concurrence ${\mathcal C}$, with
\begin{equation}\label{2-conc}
   {\mathcal C}=2|\det c|.
\end{equation}
This measure of entanglement varies between 0 (separable) and 1
(maximally entangled) and may be conveniently extended to mixed
states \cite{wootters}.

For {\em tripartite} states the situation is somewhat more
complicated. One three-dimensional analogue of the two-dimensional
determinant is the {\em Cayley hyperdeterminant}
\cite{cayley,gelfand} (denoted as Det and defined in section
(\ref{cayley})). In \cite{ckw} the Cayley hyperdeterminant (which
was termed $3-tangle$ there) is employed as a type of
hyper-concurrence to distinguish two  tripartite states: the
GHZ-state $|GHZ\rangle=1/\sqrt{2}(\ket{000}+\ket{111})$ \cite{ghz}
and the W-state
$|W\rangle=1/\sqrt{3}(\ket{001}+\ket{100}+\ket{010})$ \cite{w}.
However the Cayley hyperdeterminant {\rm Det} does not truly
reflect the nature of entanglement of these states.

For instance, the Cayley hyperdeterminant for the GHZ-state and
the W-state are one and zero, respectively. However, the W-state
{\em is} entangled; so ${\rm Det}=0$ does not provide a criterion
for separability as the simple $2 \times 2$ determinant ${\mathcal
C}$ does in the bipartite case. Further, one knows that the
W-state is in fact more robust under measurement-collapse than the
GHZ-state.  For example, if Alice measures the (first) qubit of
the GHZ-state to be $0$, then this leaves  the separable state
$\ket{00}$.  And similarly for any measurement of any qubit in any
of the three subspaces for the tripartite GHZ-state. On the other
hand, the determination of the value ``0" of any qubit in any
space for  the W-state still leaves the state (maximally)
entangled, and only if the value ``1" is measured will the
collapsed state be separable. Again this difference is not
reflected in the values of {\rm Det} for these two states. So one
needs additional indicators to reflect this difference in
entanglement properties.

In this note we supply these indicators which distinguish these
and other tripartite states, and - more significantly - provide a
necessary and sufficient criterion for the separability of a
tripartite pure state.

\section{Local Unitary Transformations}

\medskip

We initially reconsider the bipartite case. Since every (normed)
vector $v \in V$ can be transformed to the (non-entangled) state
$\ket{00}$ by a unitary transformation, it is clear that
entanglement is {\em not} invariant under unitary transformations.
However, under a {\em local unitary transformation U}, defined by
$U = U_1 \otimes U_1$, one can see that the bipartite concurrence
${\mathcal C}$ as defined in Eq.(\ref{2-conc}), for example, is
invariant:
\begin{theorem}\label{conc}
The concurrence ${\mathcal C}$ is invariant under local unitary
transformations. \newline Let $v=\sum_{i,j=1\ldots 2} a_{ij}e_i
\otimes e_j \in V=V_1 \otimes V_2$, and the unitary matrix $U=U_1
\otimes U_2$ be a local unitary matrix; then
\begin{eqnarray*}
  Uv &=& \sum c_{ij} U_1 e_i \otimes U_2 e_j\\
    &=& \sum c_{ij} (U_1)_{ik} e_k \otimes (U_2)_{jr} e_r \\
   &=& \sum c'_{kr}e_k \otimes e_r
\end{eqnarray*}where  $c'_{kr}=\sum_{ij}{c_{ij} (U_1)_{ik}(U_2)_{jr}}$
so that $c' = \tilde{U_1} c U_2$
whence
\begin{eqnarray*}
   |\det c' | &=& |\det(\tilde{U_1} c U_2)| \\
             &=& |\det{\tilde{U_1}}|\;|\det c|\;|\det{U_2}| \\
             &=&  |\det c| \; \; \;  {\rm since \; \; \; } |\det{U_i}|=1.
\end{eqnarray*}
\end{theorem}
For a bipartite  general state $\rho$ (mixed state, density
matrix) the definition of separability is:
\begin{dex}[Separable bipartite state]\label{sep-2}
The  state $\rho$ acting on $V_1 \otimes V_2$ is said to be {\em
separable} if it is given by a convex sum $\sum_{i}\lambda_i
\rho^{1}_i \otimes \rho^{2}_i\; \;  (\lambda_i \geq 0 , \; \;
\sum_{i}\lambda_{i}=1)$ where $\rho^{\alpha}_i$ acts on $
V_\alpha$.
  \end{dex}
  When $\rho=\rho^{1} \otimes \rho^{2}$ it is said to be {\em simply separable}.
  The above Definition \ref{sep-2} extends immediately to {\em multipartite} states.
\begin{dex}[Separable multipartite state]\label{sep-mult}
The  state $\rho$ acting on $V_1 \otimes V_2 \otimes \ldots
\otimes V_n$ is said to be {\em separable} if it is given by a
convex sum $\rho =\Sigma_{k}{{\lambda}_k}{\rho_1^k \otimes \rho_2^k
\otimes\ldots \otimes \rho_n^k}\; \;  (\lambda_i \geq 0 , \; \;
\sum_{i}\lambda_{i}=1)$ where $\rho^{\alpha}_i$ acts on $
V_\alpha$ .
  \end{dex}

We may see rather  immediately from Definition \ref{sep-2} that
the property of {\em being} separable is invariant under local
unitary transformations; and from Definition \ref{sep-mult}  this
extends to the multipartite case. However, an extension of the
{\it implication} from Theorem \ref{conc} that the  {\em measure}
of entanglement is preserved by local unitary transformations does
not necessarily apply to multipartite systems, since such an
extension would depend on possessing a definition of {\em
entanglement measure} for such systems, which is currently
unavailable.  Indeed, for general multipartite states, local
unitary equivalence does {\em not} preserve all the relevant
(state and substate) entanglement properties \cite{hs,hso}, as we
shall now exemplify.

\subsection{Entanglement properties of two tripartite qubit states}

\medskip

We  consider two specific examples of entangled states: the GHZ
state \cite{ghz}
\begin{equation}\label{GHZ}
|GHZ\rangle =\frac{1}{\sqrt{2}}(\ket{000}+\ket{111})
\end{equation}
and
\begin{equation}\label{cxg}
|\psi\rangle=\frac{1}{2}(\ket{100}+\ket{010}+\ket{001}+\ket{111}).
\end{equation}
In \cite{cxg} it was shown that these states are equivalent under
the local unitary transformation $U \otimes U \otimes U$ where
\begin{equation}
U=\frac{1}{\sqrt{2}}\left(
   \begin{array}{cc}
     1 & 1 \\
     -1 & 1 \\
   \end{array}
 \right).
\end{equation}
That is,
\begin{equation}\
   |\psi\rangle = U \otimes U \otimes U |GHZ\rangle.
\end{equation}

However,  the physical properties of these states are {\em not}
equivalent.  In particular, as noted in the introduction, after
qubit measurement in any subspace, the GHZ-state becomes
separable; while under a similar action the state $|\psi\rangle$
gives a maximally entangled state. Therefore any tripartite
description should distinguish between these two states.

\subsection{Cayley Hyperdeterminant}\label{cayley}

\medskip

For the tripartite qubit case we write
\begin{equation}\label{trip}
  v \in V_1\otimes V_2 \otimes V_3 =\sum_{i,j,k =1}^{2}a_{ijk} \; \; e_i \otimes e_j\otimes e_k
\end{equation}
or, as a state $\Psi$,
\begin{equation}\label{psi}
\ket{\Psi}=\sum_{i,j,k=0}^1 a_{ijk}
\ket{ijk}\; \; \; \;  (i,j,k=0,1).
\end{equation}
For $\ket{\Psi}$ the Cayley Hyperdeterminant {\rm Det} of the
coefficient hypermatrix $A=(a_{ijk})$ is defined by
\begin{dex}{Cayley Hyperdeterminant {\rm Det}}\label{Det}
\begin{eqnarray}
{\rm Det}\, A &=& a_{000}^2 a_{111}^2 + a_{001}^2 a_{110}^2 +
a_{100}^2
a_{011}^2\nonumber\\
 &-& 2\left[a_{000} a_{001} a_{110} a_{111} + a_{000} a_{010} a_{101} a_{111}
 + a_{000} a_{011} a_{100} a_{111}\right.\nonumber\\
 &+&\left.a_{001} a_{010} a_{101} a_{110}
 + a_{001} a_{011} a_{101} a_{100} +  a_{010} a_{011} a_{101} a_{100}
 \right]\label{Det-A}\\
 &+& 4\left[a_{000} a_{011} a_{101} a_{110} +  a_{001} a_{010} a_{100}
 a_{111}\right].\nonumber
\end{eqnarray}
\end{dex}

In Table \ref{tristates} we   give some examples of tripartite
qubit states.

Using  Definition \ref{Det} one may confirm by direct calculation
that ${\rm Det}=0$ for a general separable tripartite state as
given in Table \ref{tristates}.  However, ${\rm Det}=0$ for the
W-state also, which is an entangled state.  Therefore the
numerical value of {\rm Det} alone does not discriminate between
separable and non-separable states.

Further the state $|\psi\rangle$ of Eq.~(\ref{cxg}) has ${\rm
Det}=1$, as does the GHZ-state.  But, as previously noted, the
properties of retaining entanglement after qubit measurement are
completely different in the two cases.

It is clear that at the very least we need supplementary
indicators beyond the hyperdeterminant ${\rm Det}$ to specify the
entanglement properties of tripartite states, even completely
separable states.

In the following Section we propose a classification scheme.

\section{Classification}

\medskip

From the foregoing argument it would appear that one needs to
consider the subconcurrences of a tripartite state in order to
distinguish their entanglement properties, and ultimately define a
measure.

This amounts to listing the six submatrices of the hypermatrix $A$
of Eq.(\ref{Det-A}). We define
\begin{eqnarray*}\label{submat}
A_{x_0}&=& \left(a_{0ij}\right),~~
A_{x_1} =  \left(a_{1ij}\right)~~\\
A_{y_0}&=& \left(a_{i0j}\right),~~
A_{y_1} =  \left(a_{i1j}\right)~~\\
A_{z_0}&=& \left(a_{ij0}\right),~~
A_{z_1} =  \left(a_{ij1}\right).
\end{eqnarray*}
The corresponding sub-concurrences are given by the moduli of the subdeterminants\footnote{The normalization factor used here is $1$. The values in the Table are obtained by applying a normalization factor $1/|{\rm Det} A|$ for non-vanishing Det$A$ to all the terms.  For the examples given this factor is $4$. For the W-state, where ${\rm Det}A = 0$, we use the factor  $3$.}:
\begin{equation}\label{subdets}
[C_{x_0},C_{x_1},C_{y_0},C_{y_1},C_{z_0},C_{z_1}]
\end{equation}
where we have written
\begin{equation}\label{mods}
  C_{\alpha\;i} \equiv |{\rm det} A_{{\alpha}_i}| \; \; \; \; \; ({\alpha} = x,y,z;  i=0,1)
\end{equation}
which may be regarded as an {\em ordered list} that distinguishes
the bipartite substate entanglements of the given tripartite
state.

This list is by itself  not capable of discriminating between
tripartite states.  For example, it has the value $[0,0,0,0,0,0]$
for both a separable state and the GHZ state. We thus supplement
this list by the Cayley hyperdeterminant ${\rm Det}$, giving a
classification defined by the ordered list of 7 elements
\begin{equation}\label{list}
[|{\rm Det}\,A|; C_{x0}, C_{x1}, C_{y0}, C_{y1}, C_{z0}, C_{z1}].
\end{equation}

As we prove in the Appendix, the vanishing of this ordered list
provides a necessary and sufficient condition for separability,
and thus possibly paves the way to providing a useful measure for
tripartite pure qubit entanglement.

In Table \ref{tristates} we describe their classification under
the scheme herein proposed.

\begin{table}[h]
\centering \caption {\label{tristates}  Classification of some
tripartite qubit states.}
\begin{tabular}{@{}l*{15}{l}*{15}{l}}
\br
& State &  Classification\\
\mr
General Separable State & $\Sigma a_ie_i\otimes\Sigma b_je_j\otimes\Sigma c_ke_k$ & [0;0,0,0,0,0,0]\\
W-state $\ket{W}$&$1/\sqrt{3}(\ket{001}+\ket{100}+\ket{010})$ & [0;1,1,1,0,0,0]\\
GHZ-state $\ket{GHZ}$ & $(1/\sqrt{2})(\ket{000}+\ket{111})$ & [1;0,0,0,0,0,0]\\
Cluster state &
$(1/\sqrt{8})(\ket{000}+\ket{001}+\ket{100}+\ket{101}$ &
[1;1,0,1,1,0,1]\\
& $+\ket{010}-\ket{011}-\ket{110}+\ket{111})$ &\\
$\psi$-state & $(1/2)(\ket{100}+\ket{001}+\ket{010}+\ket{111})$ & [1;1,1,1,1,1,1]\\
$\phi$-State \cite{hs,hso} &$(1/2)(\ket{000}+\ket{011}+\ket{101}+\ket{110}$ & [1;1,1,1,1,1,1]\\
\br
\end{tabular}
\end{table}

\section{Discussion}

\medskip

In this note we discussed the robustness of tripartite pure qubit
states under projective measurement, and devised a classification
scheme, which consists of an ordered list of seven elements
displaying this aspect. These elements are the Cayley
hyper-determinant, and the six $2 \times 2$ subdeterminants.  In
particular we showed that this classification provides a necessary
and sufficient condition for separability.  In so far as we may
extend the definition of {\em rank} to the Cayley hyper-matrix, as
being the order of the largest non-vanishing minor, the necessary
and sufficient condition for separability may be simply stated as
that the Cayley hyper-matrix be of Rank $1$.  Further work in
progress is the extension of this definition to multipartite
systems.

\section*{Acknowledgments}
\medskip
This work is supported in part by the National Science Council
(NSC) of the Republic of China under Grant NSC
NSC-99-2112-M-032-002-MY3.

\bigskip

\appendix
\section{ Necessary and Sufficient Condition for Tripartite Qubit
Separability}

\medskip

As we argued above, to  get a better picture of the nature of
entanglement of a tripartite state it is also necessary to look at
the entanglement properties of the 2-qubits obtained when one of
the three qubits is measured.

We map the tripartite state $|\psi\rangle=\sum_{i,j,k=0}^1 a_{ijk}
|ijk\rangle$ ($i,j,k=0,1$) into the multilinear form
\begin{eqnarray}
F(x,y,z;A)=\sum_{i,j,k=0}^1 a_{ijk} \,x_iy_jz_k,~~ A=(a_{ijk}).
\end{eqnarray}
Thus the problem of factorization of $|\psi\rangle$ is reduced to
that of $F(x,y,z;A)$.  Analyzing  the entanglement of the 2-qubit
state after one qubit is measured is equivalent to determining the
factorizability of the derivatives of $F(x,y,z;A)$, namely,
\begin{eqnarray}
\frac{\partial F}{\partial z_0}&=&a_{000}x_0y_0 + a_{010}x_0y_1 + a_{100}x_1y_0 + a_{110}x_1y_1, \label{der-z0}\\
\frac{\partial F}{\partial z_1}&=&a_{001}x_0y_0 + a_{011}x_0y_1 + a_{101}x_1y_0 + a_{111}x_1y_1, \label{der-z1}\\
\frac{\partial F}{\partial y_0}&=&a_{000}x_0z_0 + a_{001}x_0z_1 + a_{100}x_1z_0 + a_{101}x_1z_1, \label{der-y0}\\
\frac{\partial F}{\partial y_1}&=&a_{010}x_0z_0 + a_{011}x_0z_1 + a_{110}x_1z_0 + a_{111}x_1z_1, \label{der-y1}\\
\frac{\partial F}{\partial x_0}&=&a_{000}y_0z_0 + a_{001}y_0z_1 + a_{010}y_1z_0 + a_{011}y_1z_1, \label{der-x0}\\
\frac{\partial F}{\partial x_1}&=&a_{100}y_0z_0 + a_{101}y_0z_1 + a_{110}y_1z_0 + a_{111}y_1z_1. \label{der-x1} 
\end{eqnarray}

For the $2\times 2\times 2$ hypermatrix $A$ in $F(x,y,z;A)$ with
components $a_{ijk}$ ($i,j,k=0,1$), the Cayley hyperdeterminant
 ${\rm Det} A$ is given as in Definition \ref{Det}.
Corresponding to the six equations (\ref{der-z0})-(\ref{der-x1}),
one defines the six determinants as in Eq.~({\ref{mods}):
\begin{eqnarray}
 C_{z_0} &=& |{\rm det} \left(a_{ij0}\right)|,~~
 C_{z_1}  =|{\rm det} \left(a_{ij1}\right)|,~~\\
 C_{y_0} &=& |{\rm det} \left(a_{i0j}\right)|,~~
 C_{y_1}  =|{\rm det} \left(a_{i1j}\right)|,~~\\
 C_{x_0} &=& |{\rm det} \left(a_{0ij}\right)|,~~
 C_{x_1}  =|{\rm det} \left(a_{1ij}\right)|.
\end{eqnarray}

We now  assert that the multilinear form $F(x,y,z;A)$, and thus
the tripartite state $|\psi\rangle$, is completely factorized,
i.e.
\begin{equation}
F(x,y,z;A)=(a_0 x_0 + a_1 x_1) (b_0 y_0 + b_1 y_1) (c_0 z_0 + c_1
z_1)\label{F-factor}
\end{equation}
for some constants $a_i, b_i$ and $c_i$,
 if and only if  the hyperdeterminant and all six
 sub-determinants are identically zero, i.e.,
\begin{eqnarray}
{\rm Det}\,A=0,~~ C_{x0}=C_{x1}=C_{y0}=C_{y1}=C_{z0}=C_{z1}=0.
\label{N-S_ Cond}
\end{eqnarray}

The necessary condition is easy to prove by direct substitution.
Below we prove the sufficient condition.

Suppose all the six sub-determinants are zero. Let us start with
$C_{z0}=C_{z1}=0$.  These conditions imply that the l.h.s. of
eqs.(\ref{der-z0}) and (\ref{der-z1}) are factorized, i.e.,
\begin{eqnarray}
&& a_{000}x_0y_0 + a_{010}x_0y_1 + a_{100}x_1y_0 + a_{110}x_1y_1\nonumber\\
 &=& (A_0 x_0 + A_1 x_1) (B_0 y_0 + B_1 y_1),\\
&& a_{001}x_0y_0 + a_{011}x_0y_1 + a_{101}x_1y_0 +
a_{111}x_1y_1\nonumber\\
 &=& (A^\prime_0 x_0 + A^\prime_1 x_1)
(B^\prime_0 y_0 + B^\prime_1 y_1)
\end{eqnarray}
for some non-zero constants $A, B, A^\prime$ and $B^\prime$. Then
from eqs.(\ref{der-z0})-(\ref{der-x1}) we have
\begin{eqnarray}
F(x,y,z;A)&=& (A_0 x_0 + A_1 x_1) (B_0 y_0 + B_1 y_1)z_0 \nonumber\\
&+& (A^\prime_0 x_0 + A^\prime_1 x_1) (B^\prime_0 y_0 + B^\prime_1
y_1)z_1
\end{eqnarray}
and
\begin{eqnarray}
a_{000}=A_0B_0,~~a_{001}=A_0^\prime B_0^\prime, ~
a_{010}=A_0B_1,~~a_{011}=A_0^\prime B_1^\prime,\\
a_{100}=A_1B_0,~~a_{101}=A_1^\prime B_0^\prime, ~
a_{110}=A_1B_1,~~a_{111}=A_1^\prime B_1^\prime,
\end{eqnarray}

For $C_{y0}=C_{y1}=C_{x0}=C_{x0}=0$, we have, respectively,
\begin{eqnarray}
\left| \begin{array}{cc} A_0B_0 & A_0^\prime B_0^\prime\\
A_1 B_0& A_1^\prime B_0^\prime
\end{array}\right|=0,~~~
\left| \begin{array}{cc} A_0B_1 & A_0^\prime B_1^\prime\\
A_1 B_1& A_1^\prime B_1^\prime
\end{array}\right|=0,\label{subdet-1}\\
\left| \begin{array}{cc} A_0B_0 & A_0^\prime B_0^\prime\\
A_0 B_1& A_0^\prime B_1^\prime
\end{array}\right|=0,~~~
\left| \begin{array}{cc} A_1B_0 & A_1^\prime B_0^\prime\\
A_1 B_1& A_1^\prime B_1^\prime
\end{array}\right|=0.\label{subdet-2}
\end{eqnarray}

For tripartite states, we have the following situations for the
solutions of eqs.(\ref{subdet-1}) and (\ref{subdet-2}):

\begin{enumerate}

\item [1.] All  $A_i, A^\prime, B_i$ and $B_i^\prime\neq 0$ ($i=0,1$):

\medskip
In this case one can factor out the common factors in each of the
four determinants in (\ref{subdet-1}) and (\ref{subdet-2}), giving
\begin{eqnarray}
\left| \begin{array}{cc} A_0& A_0^\prime\\
A_1& A_1^\prime\end{array}\right|=0,~~~
\left| \begin{array}{cc} B_0 & B_0^\prime\\
B_1& B_1^\prime
\end{array}\right|=0.\nonumber
\end{eqnarray}
Then we have
\[
\frac{A_0^\prime}{A_0}=\frac{A_1^\prime}{A_1}=p, ~~
\frac{B_0^\prime}{B_0}=\frac{B_1^\prime}{B_1}=q
\]
for some constants $p,q$, say.
 This implies $F(x,y,z;A)$ is completely
factorized
\[
F(x,y,z;A)=(A_0 x_0 + A_1 x_1) (B_0 y_0 + B_1 y_1) (z_0+pq z_1),
\]
and hence ${\rm Det} A=0$.
\bigskip

\item[2.] $X_i,~X_i^\prime\neq 0$, $X_{\bar{i}}=X_{\bar{i}}^\prime =0$ ($X=A$ or $B$, ${\bar i}=i+1 ({\rm mod}\  2)$):

\medskip
In this case , $F(x,y,z;A)$ is also factorized. To show this, let
us take $A_0,~A_0^\prime\neq 0$, and $A_1=A_1^\prime=0$. Then
$C_{y0}=C_{y1}=C_{x_1}=0$, and $C_{x0}=0$ implies
$B_0^\prime/B_0=B_1^\prime/B_1=p$ for some constant $p$.  This
implies $F(x,y,z;A)$ is factorized as
\[
F(x,y,z;A)=x_0 ( B_0y_0+B_1y_1)(A_0z_0+ p A_1 z_1),
\]
and ${\rm Det} A=0$,.

\bigskip

\item[3.] $X_i +X_i^\prime=1$,  $X_0+X_1=X^\prime_0+X^\prime_1=1$ ($X=A$ or $B$):

\medskip
There are only four cases for $X$ and $X^\prime$, namely, (i)
$A_1=B_1=0$, (ii) $A_1=B_0=0$, (iii) $A_0=B_0=0$ and (iv)
$A_0=B_1=0$. For these choices of $X$ and $X^\prime$, the six
sub-determinants are zero. This is because one of the rows or
columns of the sub-determinant is zero. Interesting cases include:
$x_0y_0z_0+x_1y_1z_1$ (GHZ-state), $x_0y_1z_0+x_1y_0z_1$,
$x_1y_1z_0+x_0y_0z_1$, and $x_1y_0z_0+x_0y_1z_1$.  They correspond
to $\{A_0,A_1,B_0,B_1\}=\{1,0,1,0\}, \{1,0,0,1\}, \{0,1,0,1\}$ and
$\{0,1,1,0\}$, respectively. The corresponding $A_i^\prime$ and
$B_i^\prime$ are obtained by making the transformations $0\to
1,~1\to 0$.

As is obvious, these states are not separable.  Thus the vanishing
of the six sub-determinants alone do not distinguish separability
of the state. Therefore the Cayley hyperdeterminant is required.
It is easy to show that in this case, ${\rm Det} A$ is given by
\begin{equation}
{\rm Det}\, A = (A_0B_0A_1^\prime B^\prime_1)^2 + (A^\prime_0
B^\prime_0A_1B_1)^2 + (A_0B_1A_1^\prime B_0^\prime)^2+ (A_0^\prime
B_1^\prime A_1B_0)^2\neq 0.
\end{equation}
Other factors in (\ref{Det-A}) are zero, as they all involve
$X_i\, X_i^\prime=0$.  The conditions $X_0+X_1\neq 0$ and
$X^\prime_0+X^\prime_1\neq 0$ guarantee that one of the four
factors in ${\rm Det} A$ is non-vanishing.

\end{enumerate}

\bigskip

Putting all these together, we see that a tripartite state is
separable iff
\[
[|{\rm Det}\,A|; C_{x0}, C_{x1}, C_{y0}, C_{y1}, C_{z0},
C_{z1}]=[0; 0,0,0,0,0,0]\]


\medskip
 \today
\end{document}